\documentclass{iopart}
\usepackage{graphicx}

\begin{document}

\title[Continuous QND feedback and unconditional spin squeezing]
{Continuous quantum nondemolition feedback and unconditional atomic spin squeezing}

\author{L K Thomsen$^\dag$, S Mancini$^\ddag$ and H M Wiseman$^\dag$}

\address{\dag\ Centre for Quantum Dynamics, School of Science,
Griffith University, Brisbane, Queensland 4111, Australia}

\address{\ddag\ INFM, Dipartimento di Fisica, Universit\`a di
Camerino, I-62032 Camerino, Italy}

%\ead{L.Thomsen@sct.gu.edu.au, Stefano.Mancini@unicam.it,
%H.Wiseman@mailbox.gu.edu.au}

\begin{abstract}
We discuss the theory and experimental considerations of a quantum
feedback scheme for producing deterministically reproducible spin
squeezing. Continuous nondemolition atom number measurement from
monitoring a probe field conditionally squeezes the sample.
Simultaneous feedback of the measurement results controls the
quantum state such that the squeezing becomes unconditional. We
find that for very strong cavity coupling and a limited number of
atoms, the theoretical squeezing approaches the Heisenberg limit.
Strong squeezing will still be produced at weaker coupling and
even in free space (thus presenting a simple experimental test for
quantum feedback). The measurement and feedback can be stopped at
any time, thereby freezing the sample with a desired amount of
squeezing.
\end{abstract}

\pacs{42.50.Dv, 32.80.-t, 42.50.Lc, 42.50.Ct}

\maketitle

\newcommand{\beq}{\begin{equation}}
\newcommand{\eeq}{\end{equation}}
\newcommand{\bqa}{\begin{eqnarray}}
\newcommand{\eqa}{\end{eqnarray}}
\newcommand{\nn}{\nonumber}
\newcommand{\nl}[1]{\nn \\ && {#1}\,}
\newcommand{\erf}[1]{Eq.~(\ref{#1})}
\newcommand{\dg}{^\dagger}
\newcommand{\smallfrac}[2]{\mbox{$\frac{#1}{#2}$}}
\newcommand{\half}{\smallfrac{1}{2}}
\newcommand{\bra}[1]{\langle{#1}|}
\newcommand{\ket}[1]{|{#1}\rangle}
\newcommand{\ip}[2]{\langle{#1}|{#2}\rangle}
\newcommand{\sch}{Schr\"odinger }
\newcommand{\schs}{Schr\"odinger's }
\newcommand{\hei}{Heisenberg }
\newcommand{\heis}{Heisenberg's }
\newcommand{\ito}{It\^o }
\newcommand{\str}{Stratonovich }
\newcommand{\dpar}[1]{\frac{\partial}{\partial {#1}}}
\newcommand{\sq}[1]{\left[ {#1} \right]}
\newcommand{\cu}[1]{\left\{ {#1} \right\}}
\newcommand{\ro}[1]{\left( {#1} \right)}
\newcommand{\an}[1]{\left\langle{#1}\right\rangle}

\section{Introduction}

Squeezed spin systems \cite{KitUed93} of atoms and ions have
attracted considerable attention in recent years due to the
potential for practical applications, such as in the fields of
quantum information \cite{QInf98} and high precision spectroscopy
\cite{Spect}. The basic principle is that the quantum correlations
of squeezed spin states will outperform classical states in the
same fashion as squeezed optical fields. Moreover, spin squeezing
is related to the fundamental concept of entanglement \cite{Bell}
and specifically represents many-particle entanglement
\cite{Soretal01,SorMol01}. There have been a number of proposals
for spin squeezing and a variety of experimental results are now
being observed.

One area of research involves the use of broadband squeezed light
\cite{sqzlight}, in which case the squeezed spin state is
generated via quantum state transfer between nonclassical light
and an atomic ensemble. This method has recently produced weakly
squeezed states \cite{Haldetal99}. In analogy with nonlinear
optics, another proposal is the collisional interactions in a
Bose-Einstein condensate (BEC). These represent a nonlinearity
which will dynamically generate spin squeezing in the trapped
state \cite{Soretal01,collBEC} and also any out-coupled beams
\cite{2beams}. There are also schemes for direct coupling to the
entangled state through intermediate states such as collective
motional modes for ions \cite{MolSor99} or molecular states for
atoms \cite{HelYou01}. A related proposal is the photodissociation
of molecular condensates \cite{PouMol01b} in analogy with the
down-conversion process in quantum optics. There has also been
experimental evidence that the ground state of a BEC confined in
an optical lattice can be produced in an atom-number squeezed
state \cite{Orzetal01}.

Production of spin squeezed states via quantum nondemolition (QND)
detection has also been considered \cite{KuzBigMan98} and spin
noise reduction using this method has been experimentally observed
\cite{KuzManBig00}. QND measurements are also involved in the
proposal \cite{Duanetal00} for the entanglement of two macroscopic
atomic samples, which has recently been achieved
\cite{JulKozPol01}. These schemes represent conditional squeezing
of the atomic ensembles. However, it would be preferable to
develop unconditional, i.e., deterministically reproducible,
squeezing especially in regard to the field of quantum
information. The ideal would be to prepare the same squeezed spin
state regardless of the measurement record.

In this work, we develop the idea introduced in
Ref.~\cite{ThoManWis02} of achieving deterministic spin squeezing
via quantum feedback, in analogy with optical squeezing via
quantum feedback \cite{WisMil94}. Monitoring the output of a probe
field sensitive to atoms in a particular internal state will
conditionally squeeze the collective angular momentum of the
sample. Unconditional squeezing is then obtained by using the
measurement results to continuously drive the system into the
desired, deterministic, squeezed spin state. We show (not done
previously) that measurement schemes based in an optical cavity or
free space both lead to the same system dynamics, i.e., that due
to a QND measurement of one collective spin operator.

It has been shown that a series of QND measurements followed by
feedback can ensure perfect agreement between the number of atoms
in two internal sates \cite{Mol99}. However this analysis does not
take into account the quantum effects of the measurement
back-action and feedback dynamics. The QND detection in our scheme
will conditionally produce quantum correlations between the atoms,
but it is not clear whether the feedback dynamics (introduced to
produce unconditional squeezing) will adversely affect these
correlations. In this paper, we perform a full quantum analysis to
look at this question, and find the quantum limit to squeezing for
the ideal situation of no atom losses. Further, we qualitatively
explore how this limit is affected by losses and what this implies
experimentally.

In Sec. 2 we briefly review the definition of squeezed spin states
and the types of nonlinear interactions which produce them. The
main body of this paper, Sec. 3, details the quantum dynamics of
the continuous measurement and feedback scheme (3.1 and 3.2) and
determines an experimentally feasible approach to implementing the
required feedback. It also includes the effective nonlinear
interactions produced by this scheme (3.3), details the numerical
work (3.4), and a discussion of experimental considerations (3.5).
Sec.~4 concludes and an appendix covers the results of inefficient
measurement.

\section{Squeezed spin systems}

The collective properties of $N$ two-level atoms are conveniently
described by a spin-$J$ system \cite{Dic54}. This is a collection
of $2J=N$ spin-\half\ particles where the internal states
$\ket{1}$ and $\ket{2}$ of the $k$th atom represent the two states
of the $k$th spin-\half\ particle. The collective angular momentum
operators, ${\mathbf J}$, are given by
$J_{\alpha}=\sum_{k=1}^{N}j^{(k)}_{\alpha}(\alpha=x,y,z)$, where
$j^{(k)}_{\alpha}=\sigma_{\alpha}^{(k)}/2$ and
$\sigma_{\alpha}^{(k)}$ are the Pauli operators for each particle.
${\mathbf J}$ obey the cyclic commutation relations
$[J_{x},J_{y}]=i\epsilon_{xyz}J_{z}$ and the corresponding
uncertainty relation $(\Delta J_{y})^{2}(\Delta
J_{z})^{2}\geq\smallfrac{1}{4}|\an{J_{x}}|^{2}$.

Equivalently, defining the operators $a\dg_{i}~(i=1,2)$ for
creating atoms in a particular internal state, we have
$J_{z}=\half(a\dg_{2}a_{2}-a\dg_{1}a_{1}),J_{+}=a\dg_{2}a_{1},J_{-}=a\dg_{1}a_{2}$
and the total number operator $a\dg_{1}a_{1}+a\dg_{2}a_{2}$ equals
a constant, i.e., $N$. $J_{z}$ represents half the total
population difference and is a quantity which can be measured, for
example by dispersive imaging techniques \cite{Andetal96}.
$J_{x}\equiv(J_{+}+J_{-})/2$ and $J_{y}\equiv(J_{+}-J_{-})/2i$
represent the two quadrature-phase amplitudes of the collective
dipole moment.

For a coherent spin state (CSS) of a spin-$J$ system, all the
elementary spins are pointing in the same mean direction. Such a
state satisfies the minimum uncertainty relationship with the
variance of the two components normal to the mean direction equal
to $J/2$. If quantum-mechanical correlations are introduced among
the atoms it is possible to reduce the fluctuations in one
direction at the expense of the other. This is the idea of a
squeezed spin state (SSS) introduced by Kitagawa and Ueda
\cite{KitUed93}, i.e. the spin system is squeezed when the
variance of one spin component normal to the mean spin vector is
smaller than the standard quantum limit (SQL) of $J/2$.

There are many ways to characterize the degree of spin squeezing
in a spin-$J$ system. We will use the criteria of S\o rensen and
co-workers \cite{Soretal01} and Wang \cite{Wang01}, where the
squeezing parameter is given by \beq
\xi^{2}_{{\mathbf{n}}_{1}}=\frac{ N(\Delta
J_{{\mathbf{n}}_{1}})^{2} } {
\an{J_{{\mathbf{n}}_{2}}}^{2}+\an{J_{{\mathbf{n}}_{3}}}^{2} },
\label{xisq} \eeq where $J_{\mathbf n}\equiv{\mathbf
n}\cdot{\mathbf J}$ and ${\mathbf n}_{i}(i=1,2,3)$ are orthogonal
unit vectors. Systems with $\xi^{2}_{\mathbf n}<1$ are spin
squeezed in the direction ${\mathbf n}$ and it has also been shown
that this indicates the atoms are in an entangled state
\cite{Soretal01,SorMol01}. This parameter also has the appealing
property that $\xi^{2}_{x}=\xi^{2}_{y}=\xi^{2}_{z}=1$ for a CSS
\cite{Wang01}.

It is well known that quadrature squeezing in optical or bosonic
systems can be generated in a nonlinear Kerr ($\chi^{(3)}$) medium.
Kitagawa and Ueda \cite{KitUed93} identified similar nonlinear
Hamiltonians that lead to squeezing in spin systems, namely
\bqa
H_{1}&=&\chi J_{z}^{2},\label{H1} \\
H_{2}&=&\chi(J_{+}^{2}-J_{-}^{2})/2i =
\chi(J_{x}J_{y}+J_{y}J_{x}), \label{H2} \eqa which can be seen as
third-order nonlinear processes if we write $\mathbf{J}$ in terms
of the atomic creation and annihilation operators. $H_{1}$ can be
thought of as twisting the $z$ axis of the quasiprobability
distribution. $H_{2}$ is then ``two-axis countertwisting'', where
the two axes bisecting the $x$ and $y$ axes are simultaneously
twisted in opposite directions. Several recent studies have shown
that $H_{1}$ describes the nonlinear spin interactions in a
two-component Bose condensate \cite{Soretal01,collBEC}.

As indicated in Ref.~\cite{KitUed93}, the variance of the squeezed
spin component, and hence \erf{xisq}, decreases to a minimum value
corresponding to the maximally squeezed spin state before
increasing again. The nonlinear interactions of Eqs.~(\ref{H1})
and (\ref{H2}) twist the quasiprobability distribution of the spin
state into, and then past, the optimally squeezed distribution.
This is because the interactions can not simply be ``turned off''
when the optimal squeezing is reached. In the limit of large
sample size, the minimum squeezing parameter for these
interactions scale, respectively, as $H_{1}:\xi^{2}\propto
N^{-2/3}$ and $H_{2}:\xi^{2}\propto N^{-1}$.

In our proposal the required nonlinear interactions between atoms
are introduced by a continuous measurement and feedback scheme. We
not only find that the minimum squeezing parameter scales as
$N^{-1}$ (in the appropriate coupling regime), but that due to the
nature of the scheme, the atomic sample could essentially be
frozen in this maximally squeezed state (or less squeezed states)
simply by ceasing the measurement and feedback.

\section{Quantum feedback scheme}

\begin{figure}
\begin{center}
\includegraphics[width=0.65\textwidth]{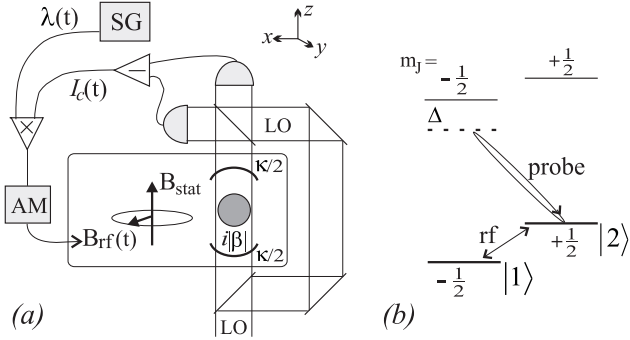}
\end{center}
\vspace{-0.5cm} \caption{\label{expt} $(a)$ Schematic experimental
configuration. A cavity field (equally damped at rate $\kappa/2$)
of amplitude $i|\beta|$ interacts with the atomic sample. The
photocurrent $I_{c}(t)$ from the homodyne detection of the cavity
output is combined with $\lambda(t)$ produced by a signal
generator (SG). The combined signal controls the amplitude (AM) of
an rf magnetic field that, together with a static field, drives
$J_{y}$. $(b)$ Single atom diagram. A static B-field lifts the
degeneracy of the magnetic sublevels. The far-detuned probe field
monitors the collective population in state $\ket{2}$ (and hence
$J_{z}$). The radio frequency (rf) driving field, applied
perpendicularly to the static field direction, induces magnetic
dipole transitions between $\ket{1}$ and $\ket{2}$ (thus driving
$J_{y}$).}
\end{figure}

Let the two internal states of each atom, $\ket{1}$ and $\ket{2}$,
be the degenerate magnetic sublevels of a $J=\half$ state, for
example, the ground state of an alkali atom. For such atoms,
transitions to the $J=\half$ excited states could then be employed
for the QND measurement, as shown in Fig.~\ref{expt}$(b)$. This
figure shows an idealized experimental schematic for our QND
measurement and feedback scheme $(a)$, and the corresponding probe
and driving transitions for each atom $(b)$. The details of this
scheme are presented in the following two sub-sections.

We assume the atomic sample is prepared such that all the atoms
are in one of their internal states and the temperature is very
low ($T\approx0$) such that we can separate out (and ignore) the
spatial degrees of freedom. A fast $\pi/2$-pulse is then applied,
coherently transferring all atoms into an equal superposition of
the two internal states, which is an eigenstate of the $J_{x}$
operator with eigenvalue $J$. As described earlier, the CSS is a
minimum uncertainty state so in this case variances of both
$J_{z}$ and $J_{y}$ are $J/2$.

\subsection{Measurement}

The feedback scheme is based on a quantum nondemolition
measurement of the atom number difference between the states
$\ket{1}$ and $\ket{2}$. The quantity to be measured is thus
represented by the operator $J_{z}$. Firstly, we consider placing
the atomic sample in a strongly driven, heavily damped, optical
cavity, as shown in Fig.~\ref{expt}$(a)$. The cavity field is
assumed to be far off resonance with respect to transitions
probing state $\ket{2}$, see Fig.~\ref{expt}$(b)$. This dispersive
interaction causes a phase shift of the cavity field proportional
to the number of atoms in $\ket{2}$. Thus, a QND measurement of
$J_{z}$ (since $N$ is conserved) is effected by the homodyne
detection of the light exiting the cavity \cite{CorMil98}.

In general, the interaction of a single atom with this far-detuned
probe field is given by the Hamiltonian \beq
V=\hbar\Omega^2(j_{z}+\half)/4\Delta \label{1atom} \eeq where
$\Omega$ is the Rabi frequency and $\Delta$ is the detuning. In
the cavity configuration with $N$ atoms this becomes
$\hbar\chi(J_z+N/2)b\dg b$, where $\chi=g^2/4\Delta$, and $g$ is
the one-photon Rabi frequency \cite{CorMil98}, and $b$,$b\dg$ are
the cavity field operators normalized such that $\an{b\dg b}$ is
the mean photon number in the cavity. Note that we can choose an
initial cavity detuning $-N\chi/2$ to eliminate the $N$ dependent
term. For strong coherent driving we can use the semiclassical
approximation $b\to i|\beta|+b$, where $b$ now represents small
quantum fluctuations around the classical amplitude $i|\beta|$.
The interaction is thus \beq H_{\rm cav} =
\hbar\chi|\beta|J_{z}(-ib+ib\dg), \label{Hcav} \eeq where we have
ignored the $\hbar\chi|\beta|^2 J_z$ term by choosing an initial
atomic detuning of $-\chi|\beta|^2$.

We assume that the cavity field is damped through both mirrors at
rate $\kappa/2$, as shown in Fig.~\ref{expt}$(a)$. The master
equation for the combined atom-field system due to the interaction
in \erf{Hcav} is thus given by \beq \dot{\rho}_{\rm
tot}=-\chi|\beta|[J_{z}(b-b\dg),\rho_{\rm tot}]+\kappa{\cal
D}[b]\rho_{\rm tot}, \eeq where ${\cal D}[r]\rho\equiv r\rho
r\dg-(r\dg r\rho + \rho r\dg r)/2$. Following the procedure of
Sec. VII in Ref.~\cite{WisMil94} we can adiabatically eliminate
the cavity dynamics if the cavity decay rate
$\kappa\gg\chi|\beta|\Delta J_{z}$, which requires
$\kappa\gg\chi|\beta|\sqrt{N}$ (since initial $\Delta
J_{z}=\sqrt{J/2}$). The evolution of the atomic system alone is
then given by \beq \dot{\rho}=M{\cal D}[J_{z}]\rho, \label{MEmeas}
\eeq where $M
=4\chi^{2}|\beta|^{2}/\kappa=8\chi^{2}P/\hbar\omega\kappa^2$ and
$P=\hbar\omega|\beta|^2\kappa/2$ is the input local oscillator
power and $\omega$ the frequency. The measurement strength $M$ is
equivalent to $2D$ in Eq.~(22) of Ref.~\cite{CorMil98}. Equation
(\ref{MEmeas}) represents decoherence of the atomic system due to
photon number fluctuations in the cavity field, with the result of
increased noise in the spin components normal to $J_{z}$.

In the alternative case of a free space QND measurement, the
analysis follows more closely that of Sec. IV A in
Ref.~\cite{ThoWis02}. From \erf{1atom}, the interaction for many
atoms in the free space limit is $\hbar\theta(J_z+N/2)p\dg p$.
Here $\theta$ is the phase shift of the probe field due to a
single atom in state $\ket{2}$, and is given by
$\theta=\hbar\omega\gamma^{2}/8A\Delta I_{\rm sat}$, where
$\omega(=2\pi c/\lambda)$ is the probe frequency, $A$ the
cross-sectional area, $\gamma$ the spontaneous emission rate and
$I_{\rm sat}= 2\pi^2\hbar\omega\gamma/\lambda^2$ for a two-level
atom \cite{Ash70}. Note that the field annihilation operators
$p,p\dg$ are normalised such that $\hbar\omega p\dg p$ is the
probe power, i.e., $p\dg p$ is the photon flux operator with units
of $s^{-1}$. The total interaction Hamiltonian (subtracting a
phase shift of $\theta/2$ for each atom) is in this case given by
\beq \label{Hfree} H_{\rm free} = \hbar\theta J_{z} p\dg p. \eeq

The back-action on the atomic sample due to this interaction can
be evaluated using the techniques of Sec. III B in
Ref.~\cite{Wis94}. Assuming the probe field is a coherent state of
amplitude $i|\varrho|$, the system evolution is \beq \dot{\rho}
=|\varrho|^{2}{\cal D}\sq{e^{-i\theta J_{z}}}\rho \simeq M {\cal
D}[J_{z}]\rho -i|\varrho|^{2}\theta[J_{z},\rho], \eeq where the
approximation requires $\theta\Delta J_{z}\ll 1$, which is
equivalent to the requirement that $\theta\sqrt{N}\ll 1$. The
measurement strength is now $M =
|\varrho|^{2}\theta^{2}=P\theta^2/\hbar\omega$, where
$P=\hbar\omega\an{p\dg p}$ is the mean power. The second term
above simply causes a frequency shift in the atomic energy level
difference, and can be ignored by choosing an initial atomic
detuning of $-|\varrho|^2\theta$.

In both cases, the evolution of the system due to this type of
interaction is given by the same expression $M{\cal
D}[J_{z}]\rho$, i.e \erf{MEmeas}. The only difference is in the
measurement strength, which is given by $M=8\chi^2
P/\hbar\omega\kappa^2$ for a cavity field and $M=\theta^2
P/\hbar\omega$ for free space. We can specifically look at the
conditioned evolution of the system due to the homodyne detection
of the probe field (this achieves the measurement of $J_z$).

Continuing the analysis for the cavity regime as in
\cite{WisMil94}, the conditioned density operator for the combined
systems evolves as \bqa d\rho_{\rm
tot}&=&\{-\chi|\beta|[J_{z}(b-b\dg),\rho_{\rm tot}]+\kappa{\cal
D}[b]\rho_{\rm tot}\}dt \nl{+}dW(t)\sqrt{\eta\kappa}{\cal
H}[b]\rho_{\rm tot}, \eqa where $dW(t)$ is an infinitesimal Wiener
increment defined by the \ito rules ${\rm E}[dW(t)]=0$;
$[dW(t)]^{2}=dt$, and ${\cal H}[r]\rho \equiv r\rho+\rho r\dg -
\Tr[(r+r\dg)\rho]\rho$. We have also introduced a measurement
efficiency $\eta$. For the setup in Fig.~\ref{expt}$(a)$, only one
cavity mirror is monitored thus giving $\eta=1/2$ (assuming
perfect photodetectors). In principle, however, one could monitor
the output of both mirrors (using, e.g., a Faraday isolator at the
input mirror) to obtain a perfect measurement, in which case
$\eta=1$. We will continue the analysis here with this assumption,
and leave the treatment of $\eta\neq1$ to the Appendix.

Adiabatic elimination as before gives the stochastic master
equation (SME) for the conditioned evolution of atomic system:
\beq d\rho_{c} = M{\cal D}[J_{z}]\rho_{c} dt + \sqrt{M}dW(t){\cal
H}[J_{z}]\rho_{c}. \label{rhoc} \eeq In this equation $\rho_{c}$
is the density matrix for the atomic system conditioned on a
particular realization of the homodyne signal given by \beq
I_{c}(t)=2\sqrt{M}\an{J_{z}}_{c}+\zeta(t) \label{current} \eeq
where $\zeta(t)=dW(t)/dt$ is a white noise term satisfying ${\rm
E}[\zeta(t)\zeta(t')]=\delta(t-t')$ and E indicates the ensemble
average. The average or unconditioned evolution is simply
recovered by averaging over all possible realizations of the
current, i.e. $\rho(t)={\rm E}[\rho_{c}(t)]$. Note that these
equations (\ref{rhoc}) and (\ref{current}) also apply for the
perfect homodyne detection of the free space field, with the
appropriate $M$.

The dominant effects of the conditioned evolution (\ref{rhoc}) on
the initial CSS are a decrease in the uncertainty of $J_{z}$
(since we are measuring $J_{z}$) with corresponding noise
increases in $J_{y}$ and $J_{x}$, and a stochastic shift of the
mean $J_{z}$ away from its initial value of zero. The latter shift
can be calculated exactly (since $d\an{J_{z}}_{c}\equiv\Tr[J_z
d\rho_c]$): \bqa
d\an{J_{z}}_{c}&=&2\sqrt{M} dW(t) (\Delta J_{z})^{2}_{c} \nn \\
&\approx& 2\sqrt{M} I_{c}(t)dt \an{J_{z}^{2}}_{c}, \label{dJzmeas}
\eqa where the approximation assumes that $\an{J_{z}}_{c}\approx0$
(such that $I_c(t)\approx dW/dt$) which will apply in the next
section. In effect, \erf{dJzmeas} is equivalent to a stochastic
rotation of the mean spin about the $y$ axis by an angle
$d\phi\approx d\an{J_{z}}_{c}/J$.

\subsection{Feedback}

\begin{figure}
\begin{center}
\includegraphics[width=0.5\textwidth]{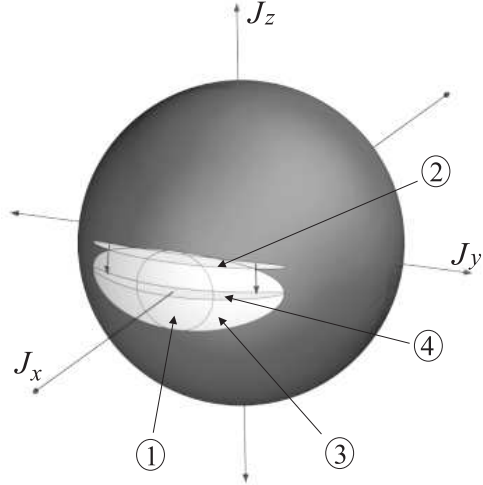}
\end{center}
\vspace{-0.5cm}\caption{\label{blobs} Schematic quasiprobability
distributions for the spin state. The spin states are represented
by ellipses on a sphere of radius $J$. The initial CSS, spin
polarized in the $x$ direction is given by state 1. State 2 is one
particular conditioned spin state after a measurement of $J_{z}$,
while state 3 is the corresponding unconditioned state due to
averaging over all possible conditioned states. The effect of the
feedback is shown by state 4: A rotation about the $y$ axis shifts
the conditioned state 2 back to $\an{J_{z}}_{c}=0$. The ensemble
average of these conditioned states will then be similar to state
4.}
\end{figure}

Monitoring the results of the measurement reduces the uncertainty
in $J_{z}$ below the SQL of $J/2$ and increases the uncertainty in
$J_{y}$, but [as shown in \erf{dJzmeas}] the mean of $J_{z}$
stochastically varies from zero. The atomic system conditioned on
a particular measurement result is thus a squeezed spin state, but
the direction of the mean spin is stochastically determined. The
unconditioned system evolution $\dot{\rho}=M{\cal D}[J_{z}]\rho$
is obtained by averaging over all the possible conditioned states,
and this leads to a spin state with $(\Delta J_{z})^{2}=J/2$, i.e.
the unmonitored measurement does not affect $J_{z}$. In other
words, the squeezed character of individual conditioned system
states is lost in the ensemble average. This is illustrated by the
states 2 and 3 in Fig.~\ref{blobs}.

To retain the reduced fluctuations of $J_{z}$ in the average
evolution we need a way of locking the conditioned mean spin
direction. This can be achieved by feeding back the measurement
results to continuously drive the system into the same squeezed
state. The idea is to cancel the stochastic shift of
$\an{J_{z}}_{c}$ due to the measurement. This simply requires a
rotation of the mean spin about the $y$ axis equal and opposite to
that caused by \erf{dJzmeas}. This is illustrated by state 4 in
Fig.~\ref{blobs}.

To make a rotation about the $y$ axis proportional to the measured
photocurrent $I_{c}(t)$, we require a Hamiltonian of the form \beq
H_{\rm fb}(t)=\hbar\lambda(t)I_{c}(t)J_{y}/\sqrt{M}=\hbar
F(t)I_{c}(t), \label{Hfb} \eeq where $F(t)=\lambda(t)J_y/\sqrt{M}$
and $\lambda(t)$ is the feedback strength. We have assumed
instantaneous feedback because that is the form required to cancel
\erf{dJzmeas}. Such a Hamiltonian can be effected, for example, by
applying either the combination of a static and an
amplitude-controlled rf magnetic field \cite{Sangetal01}, as shown
in Fig.~\ref{expt}, or two-photon excitation by
amplitude-controlled optical Raman fields. In either case, the
amplitude of the driving field is controlled by the combined
signal $\lambda(t)I_{c}(t)$ [see Fig.~\ref{expt}$(a)$].

The conditioned evolution of the system due to the feedback,
\erf{Hfb}, and measurement, \erf{rhoc}, is given by \cite{Wis94}
\beq \rho_{c}(t+dt)=e^{{\cal K}I_{c}(t)dt}\{\rho_{c}+M{\cal
D}[J_{z}]\rho_{c} dt + \sqrt{M}dW(t){\cal H}[J_{z}]\rho_{c}\},\eeq
where ${\cal K}\rho\equiv-i[F(t),\rho]$. The feedback terms of
this conditioned evolution lead to a shift in the mean $J_{z}$ by
an amount equal to \bqa d\an{J_{z}}_{\rm
fb}&=&-\lambda(t)dW(t)\an{J_{x}}_{c}/\sqrt{M}
-\lambda(t)^{2}dt\an{J_{z}}_{c}/2M
\nl{}~~~~-\lambda(t)dt\an{J_{x}J_{z}+J_{z}J_{x}}_{c} \nn \\
&\approx&-\lambda(t) I_{c}(t)dt \an{J_{x}}_{c}/\sqrt{M}.
\label{dJzfeed} \eqa As before the approximation assumes that
$\an{J_{z}}_{c}=0$ and also that the correlation between $J_{z}$
and $J_{x}$ is unchanged by the feedback. This is reasonable
because it is initially zero due to the symmetry of the CSS and
the conditioned states will remain symmetrical as shown in
Fig.~\ref{blobs}.

Since the idea is to produce $\an{J_{z}}_{c}=0$ via the feedback,
the approximations above and in \erf{dJzmeas} apply and we can
find a feedback strength such that \erf{dJzmeas} is cancelled by
\erf{dJzfeed}. The required feedback strength for our scheme is
thus \beq \lambda(t)= 2M\an{J_{z}^{2}}_{c}/\an{J_{x}}_{c},
\label{lambdac} \eeq which is obviously dependent on conditioned
averages. The feedback control, as expressed by Eqs.~(\ref{Hfb})
and (\ref{lambdac}), is essentially a form of state-estimation
based feedback and has similarities to the method of Doherty {\em
et al} \cite{Dohetal00}. Although it appears that the feedback
depends directly on the instantaneous measurement current
$I_{c}(t)$, the strength of this feedback is determined by the
conditioned state expectation values. The current only appears
directly in $H_{\rm fb}$ because of the assumptions we make about
the conditioned state.

Although \erf{lambdac} is the ideal form of the feedback strength
required to produce $\an{J_{z}}_{c}=0$, it may be not
experimentally practical since it is dependent on conditioned
expectation values. This requires classical processing of
measurement signal in real time to continually obtain the best
estimates of the system variables $J_x$ and $J_z^2$, which are
then fedback to the system via \erf{lambdac} and the Hamiltonian
(\ref{Hfb}). What would be preferable is a series of data points
or ideally an equation for $\lambda(t)$ that can be stored in a
signal generator, as shown in Fig.~\ref{expt}$(a)$, without having
to do these side calculations during the feedback process.

Applying the Markovian theory of Sec.~IVA in Ref.~\cite{Wis94}, we
obtain the continuous QND measurement and feedback master equation
for arbitrary $\lambda(t)$. It is given by \beq \dot{\rho}=M{\cal
D}[J_{z}]\rho-i\lambda(t)[J_{y},J_{z}\rho+\rho J_{z}]
+\frac{\lambda(t)^{2}}{M}{\cal D}[J_{y}]\rho. \label{ME1} \eeq The
terms in this equation describe, respectively, the measurement
back-action, the feedback driving, and the noise introduced by the
feedback. The master equation can also be rewritten in Lindblad
form \cite{Wis94}: \beq \dot{\rho}=-i\half[c\dg F(t)+F(t)c,\rho]+
{\cal D}[c-iF(t)]\rho \equiv {\cal L}\rho, \label{ME} \eeq where
the system operator $c=\sqrt{M}J_{z}$ and
$F(t)=\lambda(t)J_{y}/\sqrt{M}$ as before. Note that these
equations describe the exact unconditioned evolution of the atomic
system where the feedback strength is arbitrarily defined by
$\lambda(t)$. Equation (\ref{lambdac}) thus describes one
particular feedback scheme.

To find an experimentally suitable expression for $\lambda(t)$ we
first assume the feedback is successful in shifting the
conditioned squeezed states to $\an{J_{z}}_{c}=0$. This can be
tested by looking at the purity, ${\rm Tr}[\rho^{2}]$, of the
ensemble and conditioned states. Assuming perfect measurement, the
conditioned spin state will remain in a pure state. Thus, if the
ensemble average also has a high purity, it must comprise of
nearly identical highly pure conditioned states. Solving the
master equation, \erf{ME1}, we find that the unconditional state
does have a purity very close to one (see the numerical results
section for details), and we are justified in applying the
approximations $\an{J_{x}}_{c}\simeq\an{J_{x}}$ and
$\an{J_{z}^{2}}_{c}\simeq\an{J_{z}^{2}}$ for the feedback
strength.

We can now proceed to find analytical expressions for $\an{J_{x}}$
and $\an{J_{z}^{2}}$ and hence $\lambda(t)$. To find an expression
for $\an{J_{x}}$ we note that its decrease from $J$ is due to the
increase of $(\Delta J_{y})^{2}$ from $J/2$. The only
contributions to this increase are due to the measurement, i.e.,
the first term in \erf{ME1}, since the feedback operator ($\propto
J_{y}$) commutes with $J_{y}$. Now, for
$\an{\mathbf{J}}\sim\Or(J)$, the uncertainty in $J_{y}$ has not
increased greatly and we can use the approximation $\an{J_{x}}
\approx J\an{\cos(\Delta J_{y}/J)} \approx
J\exp[{-\an{J_{y}^{2}}/2J^{2}}]$. Here the first approximation
assumes that the angle defining the extent of the spin state in
the $xy$ plane is $\sim\Delta J_y/J$, which requires $J\gg1$, and
the second approximation assumes near Gaussian statistics for
$J_{y}$.

To find an expression for $\an{J_{z}^{2}}$ we use assume that, for
$\an{\mathbf{J}}\sim\Or(J)$, the atomic sample will approximately
remain in a minimum uncertainty state. This is equivalent to
assuming that the feedback, apart from maintaining $\an{J_{z}}=0$,
does not significantly alter the decreased variance of $J_{z}$
that was produced by the measurement. This gives
$\an{J_{z}^{2}}\approx J^{2}/4\an{J_{y}^{2}}$, where we have also
used $\an{J_{x}}\approx J$. This last step is essentially a linear
approximation represented by replacing $J_{x}$ with $J$ in the
commutator $[J_{y},J_{z}]=iJ_{x}$. From
$d_t\an{J_{y}^{2}}\equiv{\rm Tr}[J_{y}^{2}\dot{\rho}]$ we obtain
$\an{J_{y}^{2}}\approx J^{2}M t +J/2$, where the linear commutator
has been used instead of the usual cyclic commutator.

Substituting the approximation for $\an{J_{y}^{2}}$ into the
expressions for $\an{J_{x}}$ and $\an{J_{z}^{2}}$ we obtain
\bqa
\an{J_{z}^{2}} &\approx& (4M t +2/J)^{-1}, \label{appJz2} \\
\an{J_{x}} &\approx& J\exp(-M t/2), \label{appJx} \eqa and hence
for the feedback strength we have \beq \lambda(t)\approx
Me^{Mt/2}(1+NMt)^{-1}. \label{lamapp} \eeq So we now have a form
of the feedback strength which can be applied experimentally. The
question is - how well does it work?

We can analytically approximate the degree of squeezing produced
by the particular feedback scheme represented by \erf{lamapp}. For
our model
$({\mathbf{n}}_{1},{\mathbf{n}}_{2},{\mathbf{n}}_{3})=(z,x,y)$ and
the squeezing parameter, \erf{xisq}, becomes \beq
\xi^{2}_{z}=\frac{N\an{J_{z}^{2}}}{\an{J_{x}}^{2}}\simeq\frac{N
\lambda(t)}{2M \an{J_{x}}} \approx e^{Mt}(1+NMt)^{-1}.
\label{sqzapp} \eeq This leads to a minimum at $t_{*}\approx 1/M$
of \beq \xi^{2}_{\rm min}\approx e/N,~~N\gg1. \label{sqzmin} \eeq
Thus, the minimum attainable squeezing parameter asymptotically
approaches an inverse dependence on the sample size, i.e., the
Heisenberg limit. This dependence is verified numerically in Sec.
3.4.

\subsection{Effective nonlinear interaction}

The quantum correlations of our scheme can be compared with those
obtained via the different nonlinear Hamiltonians introduced
earlier, Eqs.~(\ref{H1}) and (\ref{H2}). This requires the
effective nonlinear Hamiltonian for our measurement and feedback
scheme. The master equation in Lindblad form, \erf{ME}, is by
definition separated into reversible (Hamiltonian) and
irreversible (diffusive) terms. The nonlinear spin interactions
produced in our scheme are thus given by the Hamiltonian \beq
H=\half(c\dg F(t) +F(t)c) =
\half\lambda(t)(J_{z}J_{y}+J_{y}J_{z}). \label{H} \eeq This is
equivalent to the Hamiltonian $H_{2}$ if we note that, for an
initial CSS with mean spin in the $z$ direction, \erf{H2} produces
a squeezed state in the $xy$ plane \cite{KitUed93}. Thus, the
appropriate ``two-axis countertwisting'' to squeeze the $J_{z}$
variance of of an initial CSS with mean spin in the $x$ direction,
has the form of \erf{H}. It is therefore not surprising that our
scheme, since it has essentially the same nonlinear effect on the
collective spin as $H_{2}$, also gives a $1/N$ dependence for the
minimum attainable squeezing parameter.

The advantage of our scheme is that it does not rely on internal
nonlinear spin interactions of the sample, such as those produced
by collisions in a Bose-Einstein condensate. In addition to
creating the squeezed state, these nonlinear interactions proceed
to destroy the squeezing at later times. As opposed to internal
spin interactions, which can not easily be ``switched off'', the
measurement laser and applied magnetic fields in our scheme can.
This is simply represented by setting $M=0$ in our equations,
which gives $\lambda(t)=0$ and therefore the master equation
becomes $\dot{\rho}=0$. Thus, no further evolution of the atomic
system (in the rotating frame) occurs, and the spin is essentially
``frozen'' in the squeezed state.

The exact timing for this procedure can be calculated prior to the
run if all experimental parameters are known. For
Heisenberg-limited squeezing the measurement and feedback should
be applied continuously for a time $t_*=1/M$. On the other hand,
if weaker squeezing is desired, the time is found by solving
\erf{sqzapp}. For short times and $N\gg1$ this becomes
$t\sim1/MN\xi^2$. In practice the limiting factor to squeezing
will be the time for significant atom losses, as discussed in Sec.
3.5. In any case, the degree of spin squeezing can simply be
controlled by varying the duration of the nonlinear interaction.

\subsection{Numerical results}

In this section we will justify the approximations leading up to
\erf{lamapp}, as well as verifying the $N^{-1}$ dependence of
$\xi^{2}_{\rm min}$. Regardless of the approximations, which were
only used to determine the experimentally suitable form of
$\lambda(t)$, the numerical results of this section are
\textit{exact} for the given feedback scheme. By this we mean that
the master equation (\ref{ME1}) [or \erf{ME}] is solved
\textit{exactly} for a feedback strength determined by
\erf{lamapp}. This can be done, for example, by using the {\sc
Matlab} quantum optics toolbox \cite{Tan99}.

The first approximation was that the conditioned averages could be
replaced by ensemble averages. As stated earlier, this can be
tested by looking at the purity of the unconditioned state due to
the analytical $\lambda(t)$. By iteratively solving the master
equation (\ref{ME1}) while updating $\lambda(t)$ at each time step
[using \erf{lamapp}], we find the time evolution of the purity
($=\Tr[\rho^2]$) and the expectation values ($\an{A}=\Tr[A\rho]$).
In this way the exact solution for the squeezing parameter, \beq
\xi^{2}_{z}=N\an{J_{z}^{2}}/\an{J_{x}}^{2}, \label{sqzexact} \eeq
is also found.

\begin{figure}
\begin{center}
\includegraphics[width=0.5\textwidth]{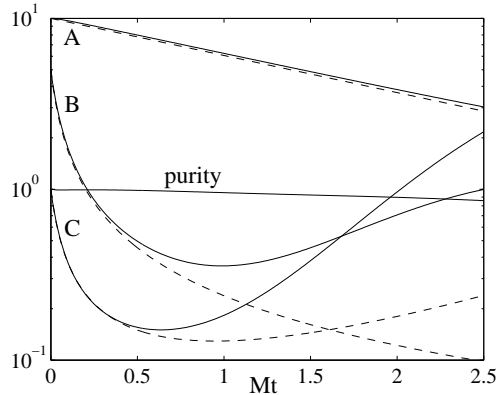}
\end{center}
\vspace{-0.5cm} \caption{\label{compare} Time dependence of the
purity $={\rm Tr}[\rho^{2}]$ (central curve), the expectation
values $\an{J_x}$ (A) and $\an{J_{z}^{2}}$ (B), and the squeezing
parameter $\xi^{2}_{z}$ (C) for $N=20$. The solid lines represent
the exact numerical solutions of the master equation with
\erf{lamapp} for the feedback strength. The dashed lines are the
corresponding analytical expressions given by Eqs.~(\ref{appJz2}),
(\ref{appJx}) and (\ref{sqzapp}). }
\end{figure}

The results of this simulation are shown in Fig. \ref{compare},
where the purity is given by the central curve, the expectation
values are given by curves A and B, and the squeezing parameter is
given by curve C. We can clearly see that the purity remains near
unity for the times of interest; at the numerical minimum of the
squeezing parameter it equals $0.978$. This implies that the
measurement and feedback scheme has worked to produce nearly
identical conditioned states. As discussed in Sec. 3.2, this
justifies the replacement of the conditioned averages of
\erf{lambdac} with ensemble averages and also the further
approximations to obtain \erf{lamapp}.

The analytical expressions for the expectation values,
Eqs.~(\ref{appJz2}) and (\ref{appJx}), and the squeezing
parameter, \erf{sqzapp}, are also included in Fig.~\ref{compare}
(dashed lines). As can be seen the analytical approximation
(dashed curve A) for $\an{J_x}$ is quite good, while the
approximation (dashed curve B) for $\an{J_z^2}$ is only good for
times up to the $\xi_z^2$ minimum. Since we are only interested in
applying the scheme up until this point, this is acceptable. As a
result though, the analytical minimum for the squeezing parameter
is not a perfect fit to the exact numerical results. However, it
is the correct order of magnitude and so we expect the analytical
scaling predicted by \erf{sqzmin}, i.e. $N^{-1}$, to be correct.

\begin{figure}
\begin{center}
\includegraphics[width=0.5\textwidth]{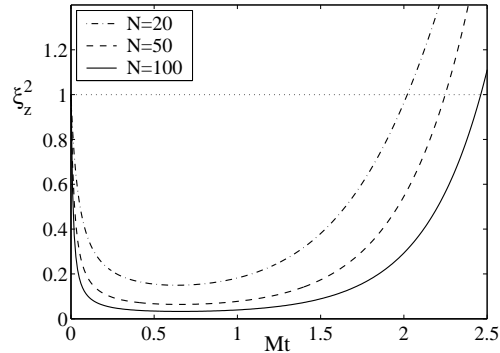}
\end{center}
\vspace{-0.5cm} \caption{\label{xifigs} Typical time dependence of
the squeezing parameter $\xi^{2}_{z}$ for the continuous
measurement and feedback scheme for three different sample sizes
$N=20$, $N=50$, and $N=100$. The dotted line represents no
measurement and no feedback, i.e., the sample remains in the CSS.}
\end{figure}

To reiterate, the analytical equation (\ref{lamapp}) for the
feedback strength succeeds in maintaining the conditional
squeezing (produced by the measurement) in a constant mean spin
direction (hence the high values for the purity of the ensemble
average state). It is also of a suitable form for experimental
realization. We now proceed to study the limit to the squeezing
produced by this scheme. In Fig.~{\ref{xifigs}} the squeezing
parameter is plotted for three different sample sizes. The dotted
line refers to the case of no measurement and no feedback. As can
be seen, the squeezing becomes stronger as the number of atoms in
the sample is increased.

\begin{figure}
\begin{center}
\includegraphics[width=0.5\textwidth]{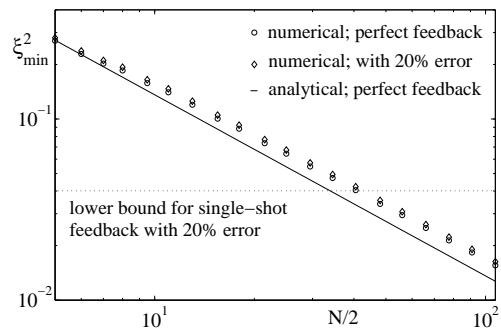}
\end{center}
\vspace{-0.5cm} \caption{\label{ximin} Atom number dependence of
the squeezing parameter minimum, $\xi^{2}_{\rm min}$. This figure
presents the results of numerical solutions of the ME, and hence
\erf{sqzexact}, with $\lambda(t)$ given by \erf{lamapp} (circles)
and $\lambda(t)\times120\%$ (diamonds). These values approach
$3.33/N$ and $3.49/N$ respectively. The dotted line is the
lower-bound for single-shot feedback also with a $20\%$ error,
while the analytical result for perfect continuous feedback $e/N$,
\erf{sqzmin}, is the solid line.}
\end{figure}

To find how the minimum of $\xi^{2}_{z}$ scales with increasing
sample size, we extract $\xi^{2}_{\rm min}$ numerically from the
above simulation. The minimum attainable squeezing parameter
approaches an inverse dependence on $N$ as the sample size
increases, as shown in Fig.~\ref{ximin}. In this figure, the
analytical result for $\xi^{2}_{\rm min}$, given by \erf{sqzmin},
is plotted along with the minima obtained from the numerical
simulation. The analytical coefficient ($e\approx2.72$) represents
an error of $\sim18\%$ compared to the numerical fit ($3.33/N$),
which is to be expected from Fig.~\ref{compare}. Nevertheless,
these errors only apply to scaling coefficients, not to the
scalings themselves. Thus, for our continuous measurement and
feedback scheme, we can conclude that the minimum attainable
squeezing parameter scales as $N^{-1}$ and occurs after a time of
the order of $M^{-1}$. The exact values for a given sample size
can be obtained by numerical simulation.

\subsection{Experimental considerations}

Our continuous scheme, as opposed to a single-shot method, is very
robust to any experimental errors in the feedback strength, as is
shown in Fig.~\ref{ximin}. The latter approach consists of a
single (integrated) measurement pulse (see e.g.,
\cite{Duanetal00}), followed by a single feedback pulse. If there
is a relative error of $\epsilon$ in the feedback strength, this
will induce an error term $(\Delta J_{z}^{\rm err})^{2}\sim
\epsilon^{2}N/4$, which will dominate the total variance for
$N\gg1$. Thus $\xi^2_{\rm min}$ will have a lower bound of
$\epsilon^{2}$, and will never scale as $N^{-1}$. On the other
hand, as shown in Fig.~\ref{ximin}, a large ($20\%$) error in
$\lambda(t)$ for continuous feedback does not affect the $N^{-1}$
scaling.

Although the analysis presented in this paper is based on
instantaneous feedback (in order to determine analytically the
limit to squeezing), including a finite delay time will have a
limited effect as long as it is of the order of $\tau_{\rm fb
}\sim(NM)^{-1}$. This is the fastest timescale in the equation
(\ref{lamapp}) for the feedback strength, and we discuss below the
implications this time has for practical application. Furthermore,
we have found the theoretical $N^{-1}$ scaling of the squeezing
parameter, $\xi_z^2$, to be unaffected by inefficient
measurements. This is detailed in the Appendix.

Experimentally, the limit to squeezing will most likely be
dominated by spontaneous atom losses due to absorption (and hence
scattering) of QND probe light. For a detailed study of the
effects of photon scattering in optical phase-shift measurements
see the recent work by Bouchoule and M\o lmer \cite{BouMol02}.
Since the scattered photons carry information about the atoms that
is not recorded, this process can also be treated in the same way
as detector inefficiency (a discussion of this is presented in the
Appendix, where we show similar scalings for $\xi^2_{\rm min}$ to
that of Ref.~\cite{BouMol02}).

In the present discussion we take the more extreme approach of
looking at atom losses from the spin-$J$ system due to absorption
of the probe light. In this way we estimate the limiting values of
the spin squeezing parameter for both cavity and free space
situations. From \erf{1atom} it can be seen that the single atom
loss rate in the far-detuned limit will be
$\Gamma=\gamma\Omega^2/4\Delta^2$, where $\gamma$ is the
spontaneous emission rate, $\Omega$ is the Rabi frequency and
$\Delta$ the detuning. For the cavity field interaction,
$\Omega^2=g^2|\beta|^2$, $M=8\chi^2P/\hbar\omega\kappa^2$ and
$\chi=g^2/4\Delta$, and so $\Gamma=\kappa\gamma M/g^2$. For the
free space model, $\Omega^2=\gamma^2I/2I_{\rm sat}$, $P=IA$, and
$M=P\theta^2/\hbar\omega$ where $\theta$ and $I_{\rm sat}$ are
defined before \erf{Hfree}, and so we obtain
$\Gamma=16\pi^2AM/\lambda^2$. The single atom loss rate can
therefore be expressed in general as \beq \Gamma=\alpha
M,\label{Lambda}\eeq where $\alpha=\kappa\gamma/g^2$ for the
cavity and $16\pi^2A/\lambda^2$ for free space, and so the number
of atoms lost by time $t$ can be estimated by \beq \Delta N=\Gamma
Nt=\alpha MNt. \label{deltaN} \eeq Therefore, by the time
($\sim1/M$) to reach the Heisenberg limit we have lost $\Delta
N\sim\alpha N$ atoms. From Sec.~3.3, the time to reach an
arbitrary amount of squeezing is $t\sim1/NM\xi^2$, so we can say
that the total number of atoms lost scales as $\Delta
N\sim\alpha/\xi^2$.

Determining the maximum number of atoms that can be lost before
destroying a given value of squeezing will thus give us some
realistic constraints that should be placed on $\alpha$. Recently,
Andr\'{e} and Lukin analyzed the squeezing produced by the
counter-twisting Hamiltonian, \erf{H2}, including the effects of
dissipation \cite{AndLuk01}. They showed that for a given amount
of spin squeezing $\xi^2=1/s$, where $1< s\leq N$, the maximum
number of atoms that can be lost without destroying the squeezing
scales as $\Delta N_{\rm max}\sim(N/s)\ln s$ \cite{AndLuk01}. In
other words, to achieve the desired level of squeezing, we need to
set $\Delta N\sim\Delta N_{\rm max}$. This becomes
$\alpha/\xi^2\sim N\xi^2\ln(1/\xi^2)$. Given an arbitrary
$\alpha$, the amount of squeezing produced including atom losses
therefore scales as \beq \xi^2\sim\sqrt{\alpha/N}. \eeq So to
produce any squeezing we need $\alpha<N$, for
$\xi^2\sim1/\sqrt{N}$ we need $\alpha\sim1$, and for the
Heisenberg limit we need $\alpha\sim1/N$.

For the case of the cavity field interaction,
$\alpha=\kappa\gamma/g^2$, so the requirement for Heisenberg
limited squeezing becomes $g^2\sim N\kappa\gamma$. This can only
be fulfilled in the very strong coupling regime, which, although
difficult, has been achieved experimentally \cite{strgexpt}.
Looking at the other regimes, we find that $\xi^2\sim1/\sqrt{N}$
squeezing will be observed for $g^{2}\sim\kappa\gamma$, and the
generation of any squeezing, i.e., $\xi^2<1$, will be observed for
$g^{2}N>\kappa\gamma$. These are the same results as found in
Ref.~\cite{AndLuk01}. As might be expected, the free space model
will not reach the Heisenberg limit. In this case
$\alpha=16\pi^2A/\lambda^2$, so the requirement becomes
$A\sim\lambda^2/16\pi^2N$ which is not possible. The limiting
value for squeezing in this case would be near $1/\sqrt{N}$, since
this would be produced if $A\sim\lambda^2/16\pi^2$, and to achieve
any squeezing we only require $A<N\lambda^2/16\pi^2$. So for the
free space regime we still expect a significant amount of
squeezing to be possible.

We can also determine the constraints that need to be placed on
the probe laser by recognizing that our analysis is based on a
far-detuned optical probe with $\Delta\gg\Omega$. This implies
that $\Gamma\ll\gamma$, and thus from \erf{Lambda},
$M\ll\gamma/\alpha$. The measurement strength for both schemes can
be re-expressed as \beq M=\gamma^2P/\hbar\omega\Delta^2\alpha^2,
\label{M}\eeq and hence our laser requirements are \beq
P/\Delta^2\ll\hbar\omega\alpha/\gamma. \eeq Take for example
$N=10^7$ Cesium atoms, where $\lambda=852$ nm ($\omega/2\pi=352$
THz) and $\gamma=5$ MHz. Then to obtain spin squeezing in either
the free space or cavity regimes, we only need
$P\ll10^{-18}\Delta^2$. On the other hand, to reach Heisenberg
limited squeezing ($\alpha\sim1/N$) we have the added requirement
that $P\ll10^{-33}\Delta^2$ as well as strong cavity coupling.
Looking at a detuning of $\Delta=1$ GHz, this restricts the laser
power to femtoWatts, which should be possible \cite{Armetal02}.

Finally, realistic feedback delay times will give some additional
restrictions on the possible squeezing. As mentioned earlier, the
feedback needs to act on a timescale of order $1/NM$, which
becomes $\tau_{\rm fb}\sim\hbar\omega\Delta^2\alpha^2/NP\gamma^2$
using the general expression for $M$. For $P=1$ fW and $\Delta=1$
GHz, this time is of the order $10\alpha^2/N$ s. Typical feedback
delay times are of order $10^{-7}$ s \cite{Armetal02}, so unless
the number of atoms is restricted to a few hundred, the Heisenberg
limit (where $\alpha\sim1/N$) will not be feasible.

During the preparation of this paper we became aware of an
analysis of the entanglement in bad cavities by S\o rensen and M\o
lmer \cite{SorMol02}. Although the scheme described in that work
is very different from ours (in fact it is similar to
Ref.~\cite{AndLuk01}), it also reaches similar results. All three
schemes require very strong cavity couplings to produce Heisenberg
limited squeezing, but weaker (and perhaps more useful) squeezing
can be produced in bad cavities and free space.

\section{Conclusion}

In conclusion we have detailed a scheme for achieving, and
controlling, the spin squeezing of an atomic sample by a
continuous quantum feedback mechanism using an optical probe. This
scheme has the advantage over previous schemes involving QND
measurement \cite{KuzBigMan98,KuzManBig00} in that it produces
unconditional, or deterministically reproducible, squeezing. This
means that the same squeezed spin state, which is also an
entangled state \cite{Soretal01,SorMol01}, will be produced
regardless of the particular measurement record. It is therefore a
more practical state for use in quantum information.

The theoretical squeezing (or entanglement) parameter approaches
the Heisenberg limit of $\xi^2\sim1/N$ for very strong cavity
couplings, a limited number of atoms, very large detuning and
small laser power. This $1/N$ scaling indicates a stronger
squeezing mechanism than collisional interactions in a
Bose-Einstein condensate where the scaling is $N^{-2/3}$
\cite{Soretal01,collBEC}. By ceasing the measurement and feedback
interactions when the desired squeezing parameter has been
reached, the squeezed state could be maintained indefinitely.
Thus, once this state is generated, there will be no further
decoherence due to the interactions of its production.

Significant squeezing will still be produced at weaker cavity
couplings, and even in free space, with correspondingly less
stringent experimental requirements. The free space QND
measurement regime therefore presents a relatively simple, and
interesting, experimental implementation of quantum feedback, an
area which is still in its infancy. Also, the weaker the squeezing
the less vulnerable the state is to further decoherence after its
production. The Heisenberg-limited squeezed state will be severely
affected by the loss of a few atoms, whereas less strongly
squeezed states can lose many without serious damage
\cite{AndLuk01}. The states produced in a free space scheme might
therefore be adequate for practical quantum information, where
other experimental factors could destroy the more sensitive
states. Then, if necessary, one can devise entanglement
purification procedures for atomic samples \cite{Dunetal02}.
Finally, achieving and controlling a spin squeezed state could
also be useful in obtaining tomographic phase space pictures of
nonclassical states of atomic samples \cite{tomoBEC}.

\appendix

\section{Inefficient measurements}

In real experiments there is a finite efficiency involved in the
measurement process. We now include this in our analysis to see
whether the degree of spin squeezing is significantly affected.
Basically, only a certain proportion ($\eta$) of the probe beam is
actually detected and then used for feedback, but the undetected
proportion ($1-\eta$) still causes back-action on the system. This
could be due to only monitoring the output of one side of a cavity
[as in Fig.~\ref{expt}$(a)$], or imperfect photodetectors, or even
undetected photons scattered off the atomic sample (which is
treated more rigorously in Ref.~\cite{BouMol02}).

Let us first look at the effect of an inefficient measurement of
$J_{z}$, before adding the feedback. The stochastic master
equation $d\rho_{c}$ is now made up of two contributions: the
actual homodyne detection of a proportion $\eta$ of the probe
light \beq \eta{\cal D}[c]\rho_{c} dt + \sqrt{\eta}dW(t){\cal
H}[c]\rho_{c}, \label{rhoceta} \eeq plus the back-action due to
the probe light that wasn't detected \beq (1-\eta){\cal
D}[c]\rho_{c} dt, \eeq where $c=\sqrt{M}J_{z}$ as before.
Similarly, the homodyne photocurrent is now given by \beq
I_{c}(t)=\eta\an{c+c\dg}_{c}+\sqrt{\eta}\xi(t). \label{Iineff}
\eeq The stochastic shift in $\an{J_{z}}_{c}$ due to an
inefficient measurement is therefore $2\sqrt{\eta M}dW(t)(\Delta
J_{z})^{2}_{c}$. If we again assume $\an{J_z}_c\approx0$ (which
will be true when the feedback is included) then
$I_c(t)\approx\sqrt{\eta}dW/dt$ and so we again obtain the
approximation in \erf{dJzmeas}.

Including feedback to cancel this stochastic shift, we obtain the
same feedback Hamiltonian (\ref{Hfb}) as before, although note
that the current $I_{c}(t)$ is now defined by \erf{Iineff}. Thus,
the shift in the mean $J_{z}$ due to feedback is still given by
\erf{dJzfeed} and the optimal feedback strength is therefore also
unchanged from the form given in \erf{lambdac}. It is the
assumptions made for obtaining the analytical expression for
$\lambda(t)$ that are now complicated.

The master equation for the inefficient measurement and feedback
is given by \beq \dot{\rho}=(1-\eta+\eta){\cal
D}[c]\rho-i[F,c\rho+\rho c\dg] +\frac{1}{\eta}{\cal D}[F]\rho,
\eeq where $c=\sqrt{M}J_z$ and $F=\lambda(t)J_{y}/\sqrt{M}$ as
before. Thus, the total back-action due to the probe does not
depend on the detection efficiency since it is made up of both the
detected ($\eta$) and undetected ($1-\eta$) contributions. The
Lindblad form of the master equation, \erf{ME}, now also has the
additional term $[(1-\eta)/\eta]{\cal D}[F]\rho$.

To simplify the form of $\lambda(t)$ in \erf{lambdac} we again
assume the feedback is successful and replace the conditioned
averages with ensemble averages. To find $\an{J_{z}^{2}}$ we
earlier assumed a near minimum uncertainty state, but this cannot
be true for imperfect measurements. To use $\an{J_{z}^{2}}\approx
J^{2}/4\an{J_{y}^{2}}$ for $\eta<1$ we must therefore calculate
the average $J_{y}^{2}$ only using the evolution due to the actual
measurement and feedback, i.e. we ignore $(1-\eta){\cal
D}[c]\rho$. This gives $\an{J_{y}^{2}}\approx J^{2}\eta M t+J/2$
and then $\an{J_{z}^{2}}\approx(4\eta M t+2/J)^{-1}$. To find
$\an{J_{x}}$ we earlier noted that it decreased indirectly due to
the measurement back-action. As shown above, the total back-action
due to the probe beam is independent of detection efficiency and
so we can keep the earlier approximation (\ref{appJx}) for
$\an{J_{x}}$.

These approximations for the averages this time lead to a feedback
strength of \beq \lambda(t) \approx Me^{Mt/2}(1+\eta NMt)^{-1}.
\label{lamappeta} \eeq The approximate squeezing parameter then
becomes \beq \xi^{2}_{z}\approx e^{Mt}(1+\eta NMt)^{-1}, \eeq
which has a minimum at $t_*\approx 1/\eta M$ of \beq \xi^{2}_{\rm
min}\approx e^{1/\eta}/N,~~N\gg1. \label{sqzmineta} \eeq So for
$\eta<1$, the minimum attainable squeezing parameter is greater
than that for $\eta=1$. However, finite efficiency only affects
the coefficient of this minimum - the $N^{-1}$ dependence is
unaffected. Note that this is the same scaling with the number of
atoms as found in Eq.~(59) of Ref.~\cite{BouMol02}, which confirms
that photon scattering can also be regarded as a measurement
inefficiency.

\end{document}